\documentclass[10pt,conference]{IEEEtran}
\usepackage[cmex10]{amsmath}

\newtheorem{definition}{Definition}

\newcommand{\ie}{\emph{i.e.}, }
\newcommand{\eg}{\emph{e.g.}, }
\newcommand{\pnt}[1]{{\mbox{\boldmath $#1$}}}
\newcommand{\cof}[2]{\mbox{$#1_{\boldsymbol{#2}}$}}

\newcommand{\V}[1]{\mbox{$\mathit{Vars}(#1)$}}
\newcommand{\Va}[1]{\mbox{$\mathit{Vars}(\boldsymbol{#1})$}}

\newcommand{\s}[1]{\mbox{$\{#1\}$}}

\newcommand{\nGz}[2]{$G_{non-\{z\}}$}

\newcommand{\prr}[1]{\mi{Prev}(\boldsymbol{q})}

\newcommand{\mi}[1]{\mathit{#1}}
\newcommand{\ti}[1]{\textit{#1}}
\newcommand{\tb}[1]{\textbf{#1}}

\newcommand{\Ds}[2]{\mbox{\pnt{#1}~$\rightarrow$ \s{#2}}}

\newcommand{\ttt}{\>\>\>}

\newcommand{\Tt}{\>\>}
\newcommand{\Sub}[2]{\mbox{$\mi{#1}_\mi{#2}$}}

\newcommand{\prob}[2]{\mbox{$\exists{#1} [#2]$}}

\newcommand{\DS}{\mbox{$\Omega$}}

\newcommand{\Cdi}{\ti{DCDS}~}
\newcommand{\ecnf}{\ensuremath{\exists\mathrm{CNF}}}
\newcommand{\Comment}[1]{}
\newcommand{\Dss}[4]{\mbox{$(\prob{#1}{#2},\pnt{#3})~\rightarrow #4$}}
\newcommand{\DDs}[4]{\mbox{$(\prob{#1}{#2},\pnt{#3})~\rightarrow$ \s{#4}}}
\newcommand{\PQE}{\ti{DS-PQE}~}
\newcommand{\pqe}{\ti{DS-PQE}}
\newcommand{\pri}{\Sub{PR}{init}~}
\newcommand{\prf}{\mbox{$\mi{PR}(\pnt{q})$}~}
\newcommand{\Pqe}{\ti{DS\_PQE}~}
\newcommand{\mcq}{\ti{MC-QE}~}
\newcommand{\mcpq}{\ti{MC-PQE}~}

\newcommand{\Impl}[2]{\mbox{$#1 \rightarrow #2$}}
\newcommand{\Nmpl}[2]{\mbox{$#1 \not\rightarrow #2$}}
\newcommand{\sat}{\ti{PQE-SAT}}
\newcommand{\SAT}{\ti{PQE-SAT}~}

\usepackage{wrapfig}
\usepackage{graphicx}
\IEEEoverridecommandlockouts
\begin{document}

\title{Partial Quantifier Elimination}

\author{\IEEEauthorblockN{Eugene Goldberg and Panagiotis Manolios} 
\IEEEauthorblockA{
Northeastern University, USA, 
 \{eigold,pete\}@ccs.neu.edu}}

\maketitle

\begin{abstract}
We consider the problem of Partial Quantifier Elimination
(PQE)\footnote{The only difference of this technical report from the previous
version~\cite{first_version} is as follows. The description of the
algorithm given in~\cite{first_version} was missing a case.  (The
implementation that we tested in experiments was correct but the
pseudo-code of the algorithm we gave missed a few lines addressing the
case in question.) The missing part of the algorithm is described in
Section~\ref{sec:algorithm} of this report.
}.  Given formula \prob{X}{F(X,Y)
  \wedge G(X,Y)}, where $F, G$ are in conjunctive normal form, the PQE
problem is to find a formula $F^*(Y)$ such that $F^* \wedge
\prob{X}{G} \equiv \prob{X}{F\wedge G}$.  We solve the PQE problem by
generating and adding to $F^*$ clauses over the free variables that
make the clauses of $F$ with quantified variables
\emph{redundant}. The traditional Quantifier Elimination problem (QE)
is a special case of PQE where $G$ is empty so \ti{all} clauses of the
input formula with quantified variables need to be made redundant.
The importance of PQE is twofold. First, many problems are more
naturally formulated in terms of PQE rather than QE.  Second, in many
cases PQE can be solved more efficiently than QE.  We describe a PQE
algorithm based on the machinery of dependency sequents and give
experimental results showing the promise of PQE.
\end{abstract}

\section{Introduction}
\label{sec:introduction}
The elimination of existential quantifiers is an important problem
arising in many practical applications.  We will refer to this
problem as the Quantifier Elimination problem, or QE. Given a formula
\prob{X}{F} where $F$ is a propositional formula, the \tb{QE
  problem} is to find a quantifier free formula $G$ such that $G
\equiv \prob{X}{F}$. In this paper, we assume that all
propositional formulas are represented in conjunctive normal form
(CNF).

Unfortunately, the efficiency of current QE algorithms still
leaves much to be desired.  This is one reason that many
successful theorem proving methods such as interpolation and IC3
avoid QE and use SAT-based reasoning instead.  These methods can
be viewed as solving specialized versions of the QE problem that
can be solved efficiently. For example, finding an interpolant
$I(Y)$ of formula $A(X,Y) \wedge B(Y,Z)$ comes down to solving a
special case of QE where $I \equiv \prob{X}{A}$ needs to hold
only in subspaces where $B \equiv 1$.  So it is important to perform a
systematic study of the QE problem, looking for variants of the
problem that can be solved efficiently.  Such a study can help us
better understand existing algorithms that sidestep the use of QE
in favor for more limited, specialized methods. The study may
also lead to the discovery of new applications of QE.

In this paper, we consider a variation of the QE problem called
Partial QE (PQE). Let \prob{X}{F(X,Y)\wedge G(X,Y)} be a formula
where variables of $X$ are quantified.  The \tb{PQE problem} is
to find a formula $F^*(Y)$ such that $F^* \wedge \prob{X}{G}
\equiv \prob{X}{F\wedge G}$.  We will say that $F^*$ is obtained
by \tb{taking} \pnt{F} \tb{out of the scope of the quantifiers}.
Note that if $F^* \rightarrow \prob{X}{G}$ holds, then $F^*
\equiv \prob{X}{F \wedge G}$. That is, in this case, a solution
to the PQE problem is also a solution to the QE problem. We will
say that in this case QE reduces to PQE.

Our motivation for solving the PQE problem is twofold. First, in
many cases, a verification problem can be formulated as an
instance of PQE rather than QE. Besides, even if the original
problem is formulated in terms QE it can sometimes be reduced to
PQE. Second, in many cases, the PQE
problem can be solved much more efficiently than QE. We are
especially interested in applying PQE when formula $F$ is much
smaller than $G$.

The relation between efficiency of solving PQE and QE can be
better understood in terms of clause redundancy~\cite{fmcad13}.
The PQE problem specified by \prob{X}{F \wedge G} reduces to
finding a set of clauses $F^*$ that makes all $X$-clauses of $F$
redundant in formula \prob{X}{F \wedge G}.  (An
\pnt{X}\tb{-clause} is a clause that contains a variable from
$X$.) Then every clause of $F$ can be either dropped as redundant
or removed from the scope of the quantifiers as it contains only
free variables.

One can view the process of building $F^*$ as
follows. $X$-clauses of $F$ are made redundant in \prob{X}{F
  \wedge G} by adding to $F$ resolvent clauses derived from $F
\wedge G$.  Notice that no clause obtained by resolving \ti{only}
clauses of $G$ needs to be made redundant. Adding resolvents to
$F$ goes on until all $X$-clauses of the current formula $F$ are
redundant.  At this point, the $X$-clauses of $F$ can be dropped
and the remaining clauses of $F$ form $F^*$.

If $F$ is much smaller than $G$, the process of solving PQE looks like
wave propagation where $F$ is the original ``perturbation'' and $G$ is
the ``media'' where this wave propagates. Such propagation can be
efficient even if $G$ is large. By contrast, when solving the QE
problem for \prob{X}{F \wedge G} one needs to make redundant the
$X$-clauses of both $F$ and $G$ and \ti{all} resolvent $X$-clauses
including the ones obtained by resolving only clauses of $G$.

In this paper, we describe a PQE-algorithm called \PQE that is based
on the machinery of D-Sequents~\cite{fmcad12,fmcad13}.  One needs this
machinery for PQE for the same reason as for QE~\cite{fmcad12}.  Every
clause of $F^*(Y)$ can be obtained by resolving clauses of $F \wedge
G$.  However, the number of clauses that are implied by $F \wedge G$
and depend only on $Y$ is, in general, exponential in $|Y|$. So it is
crucial to identify the moment when the set of clauses derived so far
that depend only on $Y$ is sufficient to make the $X$-clauses of $F$
redundant in \prob{X}{F \wedge G}.  The machinery of D-sequents is
used for such identification. Namely, one can stop generating new
clauses when a D-sequent stating redundancy of the $X$-clauses of $F$
is derived. We experimentally compare \PQE with our QE algorithm
from~\cite{fmcad13} in the context of model checking.

The following exposition is structured as follows. In
Sections~\ref{sec:mod_check} and~\ref{sec:sat}, we discuss some
problems that can benefit from an efficient PQE-algorithm.  A run of
\PQE on a simple formula is described in Section~\ref{sec:example}.
Sections~\ref{sec:basic_defs} and~\ref{sec:dseqs} give basic
definitions and recall the notion of D-Sequents. In
Section~\ref{sec:algorithm}, \PQE is described. We discuss previous
work in Section~\ref{sec:background}.  Experimental results are given
in Section~\ref{sec:experiments}. Finally, we make conclusions in
Section~\ref{sec:conclusions}.

\section{Using PQE For Model Checking}
\label{sec:mod_check}
In this section and the one that follows we describe some applications where using an
efficient PQE solver can be very beneficial. 
A few more applications of PQE are listed in 
an extended abstract~\cite{south_korea}.
\label{sec:mc}

%
%
\subsection{Computing pre-image in backward model checking}
\label{subsec:pre-image}
Let $T(S,S')$ be a transition relation where $S$ and $S'$ specify
the current and next state variables respectively.  We will refer
to complete assignments \pnt{s} and \pnt{s'} to variables $S$ and
$S'$ as present and next states respectively.  Let formula
$H(S')$ specify a set of next-states and $G(S)$ specify the
pre-image of $H(S')$.  That is, a present state \pnt{s} satisfies
$G$ iff there exists a next state \pnt{s'} such that $H(\pnt{s'})
\wedge T(\pnt{s},\pnt{s'}) = 1$.

Finding $G$ reduces to QE that is to building a formula logically
equivalent to \prob{S'}{H \wedge T}.  However, one can construct
the pre-image of $H$ by PQE as follows.  Let $H^*$ be a formula
such that $H^* \wedge \prob{S'}{T} \equiv \prob{S'}{H \wedge T}$
\ie $H^*$ is a solution to the PQE problem.  Notice that $H^*$
implies \prob{S'}{T} because $\prob{S'}{T} \equiv 1$. Indeed, for
every present state \pnt{s} there always exists some next state
\pnt{s'} such that $T(\pnt{s},\pnt{s'}) = 1$. So $H^* \equiv
\prob{S'}{H \wedge T}$ and hence specifies the pre-image of
$H$. In other words, here QE reduces to PQE.

%
%
\subsection{State elimination in IC3-like model checkers}
\label{subsec:ic3}
In this subsection, we discuss state elimination, 
a key problem for IC3-like model checkers~\cite{ic3}.  Given a
transition relation $T(S,S')$, the problem of eliminating a state
\pnt{s} is to find a clause $C$ falsified by \pnt{s} and
inductive relative to a formula $F$. The latter means that 
$F \wedge C(S) \wedge T \rightarrow C(S')$.

The performance of IC3 strongly depends on the efficiency of
solving the state elimination problem and the quality of
inductive clauses generated to solve it. An IC3-like model
checker would benefit from an efficient algorithm finding the
pre-image of the state \pnt{s} to be eliminated~\cite{sasha}.
Finding the pre-image of \pnt{s} can be useful when no inductive
clause $C$ eliminates \pnt{s}. In this case, IC3 removes some
states that satisfy $F$ and from which a direct transition to
\pnt{s} is possible. This is done by adding new clauses to $F$,
which eventually leads to appearance of a clause $C$ that is
inductive relative to $F$ and eliminates \pnt{s}. Finding the
best states to remove is crucial for the performance of IC3.  The
pre-image of \pnt{s} can be very useful to identify such states.

Finding the pre-image of \pnt{s} is a special case of the problem
we discussed in Subsection~\ref{subsec:pre-image}.  Let $H$ be
the set of unit clauses specifying state \pnt{s} \ie \pnt{s}
satisfies $H$.  Let $G(S)$ be a formula such that $F \wedge G
\wedge \prob{S'}{T} \equiv F \wedge \prob{S'}{H(S') \wedge T}$.
The complete assignments satisfying $G$ specify the pre-image of
\pnt{s} ``relative'' to $F$.  Any clause $C$ inductive
relative to $F$ has to be falsified by assignments satisfying
$F \wedge G$.  The PQE-algorithm we describe in this paper is
not efficient enough to be used in the loop of IC3 right
away, but this may change soon.

%
%
\section{Using PQE For SAT-solving}
\label{sec:sat}
In this section, we describe a SAT-algorithm based on PQE.  (We
will refer to this algorithm as \sat.) We also contrast \SAT with a
SAT-solver based on Conflict Driven Clause Learning (CDCL).
%
%
\subsection{High-level view of the algorithm}
 The pseudocode
of \SAT is shown in Figure~\ref{fig:sat}. Let $G(X)$ be a CNF formula
to be checked for satisfiability. In the main loop, \SAT performs the
following actions.  First, it generates a clause $C$ that is not
trivially subsumed by a clause of $G$ (line 2).  Then \SAT solves an
instance of the PQE problem (line 3). Namely, it calls procedure
\ti{SolvePQE} to find formula $R$ such that $R \wedge \prob{X}{G}
\equiv \prob{X}{C \wedge G}$. Depending on the type of formula $R$
returned by \ti{SolvePQE}, \SAT either updates $G$ by adding a clause
or makes a final decision on whether $G$ is satisfiable (lines 4-12).

%
%
\setlength{\intextsep}{4pt}
\setlength{\textfloatsep}{10pt}
\begin{figure}
\small
\vspace{5pt}
\begin{tabbing}
aaa\=bb\= cc\= ddddddd\= \kill
$\mi{SAT\_by\_PQE}(G)$\{\\
\tb{\scriptsize{1}}\> while ($\mi{true}$) \{\\
\tb{\scriptsize{2}}\Tt $C := \mi{GenClause}(G)$;\\
\tb{\scriptsize{3}}\Tt$R := \mi{SolvePQE}(\prob{X}{C \wedge G})$;\\  
\> - - - - - - - - - - - -  \\
\tb{\scriptsize{4}}\Tt if ($R$ is derived without using $C$) \{\\
\tb{\scriptsize{5}}\ttt   $G := \mi{AddClause}(G,R)$;\\
\tb{\scriptsize{6}}\ttt   if ($R \equiv 0$) return($\mi{UNSAT}$);\\
\tb{\scriptsize{7}}\ttt   continue;\}  \\
\> - - - - - - - - - - - -  \\
\tb{\scriptsize{8}}\Tt if ($R \equiv 1$) \{\\  
\tb{\scriptsize{9}}\ttt  $G : = G \cup \s{C}$; \\
\tb{\scriptsize{10}}\ttt  continue; \}\\  
\> - - - - - - - - - - - -  \\
// the only possibility left is $R \equiv 0$ \\
\tb{\scriptsize{11}}\Tt if (\Impl{G}{C}) return($\mi{UNSAT}$);  \\  
\tb{\scriptsize{12}}\Tt else return($\mi{SAT}$);   \\
\end{tabbing} 
\vspace{-20pt}
\caption{SAT checking by PQE}
\label{fig:sat}
\end{figure}

\ti{SolvePQE} returns three kinds of formula $R$. The actions \SAT
take for every kind of formula $R$ are separated by the dotted lines
in Figure~\ref{fig:sat}. We will refer to a formula $R$ returned by
\ti{SolvePQE} as a formula of the \ti{first kind} if it is obtained by
resolving only clauses of $G$ (lines 4-7).  In this case, $R$ is just
a clause that subsumes $C$. (In particular, $R$ can be equal to $C$.)
On the one hand, the fact that $R$ is derived without using clause $C$
means that $R$ is implied by $G$. On the other hand, the fact that $R$
subsumes $C$ suggests that $C$ is also implied by $G$. Thus $C$ is
trivially redundant in \prob{X}{C \wedge G}. \SAT adds clause $R$ to
$G$. If clause $R$ is empty, then $G$ is obviously unsatisfiable.

If resolution derivation of the formula $R$ returned by \ti{SolvePQE} involves
clause $C$ we will refer to $R$ as a formula of  the second or third kind.
In this case, $R$ is a constant. That is $R$
either has no clauses (formula of the \ti{second kind}, $R \equiv 1$) or it
is an empty clause (formula of the \ti{third find}, $R \equiv 0$).  Indeed,
just derivation of a clause $A$ subsuming $C$ does not mean that $C$
is redundant in \prob{X}{C \wedge G}. The reason is that $A$ is
derived using clause $C$ and so $A$ may not be implied by $G$. On the
other hand, if $A$ is not empty (and hence contains variables of $X$) ,
it cannot be taken out of the scope of quantifiers.

Actions of \SAT when \ti{SolvePQE} returns a formula $R$ of the second
kind are shown in lines 8-10.  The fact that $R \equiv 1$ means that
$C$ is redundant in \prob{X}{C \wedge G}. That is either $C$ is
implied by $G$ or $C$ eliminates some (but not all) assignments
satisfying $G$. In either case, $C \wedge G$ is equisatisfiable to
$G$. For that reason \SAT adds $C$ to $G$.

What \SAT does when \ti{SolvePQE} returns a formula $R$ of the third
kind is shown in lines 11-12. The fact that $R \equiv 0$ means that
either $G$ is unsatisfiable or $C$ is falsified by \ti{every}
assignment satisfying $G$.  \SAT tells these two cases apart by
checking if $C$ is implied by $G$.
%
%
\subsection{Difference between \SAT and a CDCL SAT-solver}
The difference between \SAT and a CDCL SAT-solver is twofold. First,
\SAT employs non-resolution derivation of clauses. This derivation
occurs, when \ti{SolvePQE} returns a formula $R$ of the second kind
(i.e. $R \equiv 1$).  In contrast to a formula of the first kind, in
this case, \ti{SolvePQE} proves that $C$ is redundant in \prob{X}{C
  \wedge G} without generation of a clause subsuming $C$. A simple
example of a clause obtained by non-resolution derivation is a blocked
clause ~\cite{blocked_clause} (see Section~\ref{sec:example}). Adding
clauses obtained by non-resolution derivation allows one to get proofs
that are much shorter than those based on pure resolution.  For
example, in~\cite{gen_ext_resol} it was shown that extending
resolution with a rule allowing to add blocked clauses makes it
exponentially more powerful.

The second difference between \SAT and a CDCL SAT-solver is in the way
they generate a satisfying assignment. When \ti{SolvePQE} returns an
empty clause (a formula of the third kind) it checks if $G$ implies
$C$.  A counterexample showing that \Nmpl{G}{C} is also an assignment
satisfying $G$. Checking if \Impl{G}{C} holds reduces to testing the
satisfiability of $G$ in the subspace where $C$ is falsified.

As far as finding a satisfying assignment is concerned, \SAT
potentially has three advantages over CDCL-solvers. The first advantage
is that \SAT can derive clauses that eliminate satisfying assignments
of $G$.  This is important because the ability of a CDCL-solver to
efficiently find a satisfying assignment hinges on its ability to
derive short clauses.  For example, if a unit clause $\overline{v}$ is
derived by a CDCL-solver, it can immediately set $v$ to 0. However,
such a clause cannot be derived if formula $G$ has satisfying
assignments with $v=0$ and $v=1$.  The ability of \SAT to add clauses
removing satisfying assignments in general leads to enhancing the
quality of learned clauses. Suppose, for example, that \SAT adds to
$G$ a clause $C$ that eliminates all satisfying assignments with $v=1$
(but preserves at least one satisfying assignment with $v=0$). Then
formula $G$ implies clause $\overline{v}$ and hence the latter can be
derived from $G$ by resolution.

The second advantage of \SAT is that if clause $C$ is long (i.e. $C$
has many literals), then checking \Impl{G}{C} can be much simpler than
just testing the satisfiability of $G$. The third advantage of \SAT is
that in case $C$ is short \SAT can exploit the resolution derivation
of an empty clause it obtained. Let $P$ denote such a derivation
produced by \sat.  The fact that $G$ is satisfiable and $C \wedge G$
is not means that every assignment satisfying $G$ falsifies $C$. This
entails that every cut of $P$ must contain either clause $C$ itself or
a descendant clause $A$ of $C$ such that \Nmpl{G}{A}. Note that even
if $C$ is a short clause, it can have descendants that are very long.
So if $C$ is short, one can replace computationally hard check
\Impl{G}{C} with a sequence of checks \Impl{G}{A} starting with the
longest descendant clauses of $C$.

\section{Example}
\label{sec:example}
In this section, we describe a run of a PQE algorithm called \PQE
that is described in Section~\ref{sec:algorithm}.  \PQE is a
modification of the QE algorithm called \Cdi\cite{fmcad13} based
on the machinery of Dependency sequents (D-sequents).  In this
section, we use notions (\eg that of D-sequents) that will be
formally defined in Section~\ref{sec:dseqs}.  Recall that an
$X$-clause is a clause that contains at least one variable from a
set $X$ of Boolean variables.

Let $F=C_1 \wedge C_2$ where $C_1 = y \vee x_1$, $C_2 =
\overline{y} \vee x_3$ Let $G = C_3 \wedge C_4 \wedge C_5 \wedge
C_6$ where $C_3 = \overline{x}_1 \vee x_2$, $C_4 = \overline{x}_1
\vee \overline{x}_2$, $C_5 = \overline{x}_3 \vee x_4$, $C_6 = y
\vee \overline{x}_4$.  Let $X=\s{x_1,x_2,x_3,x_4}$ be the set of
variables quantified in formula \prob{X}{F \wedge G}. So $y$ is
the only free variable of $F \wedge G$.

\ti{Problem formulation.} Suppose one needs to solve the PQE
problem of taking $F$ out of the scope of the quantifiers in
\prob{X}{F \wedge G}. That is one needs to find $F^*(y)$ such
that $F^* \wedge \prob{X}{G} \equiv \prob{X}{F \wedge G}$.
Below, we describe a run of \PQE when solving this problem.

\setlength{\intextsep}{8pt}
\begin{wrapfigure}{l}{1.6in}
 \begin{center}
 \includegraphics[height=30mm]{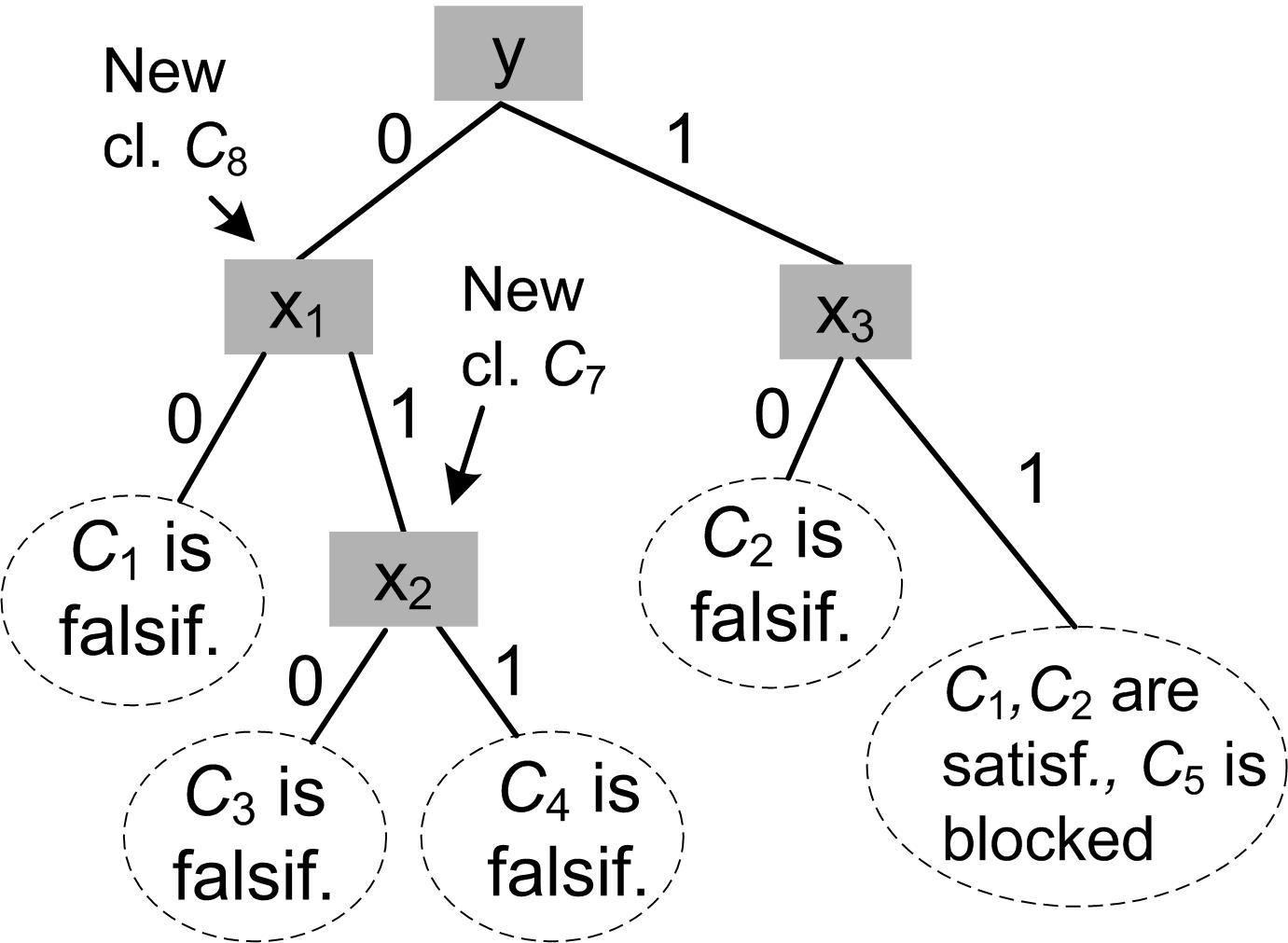}
  \end{center}
\vspace{-3pt}
\caption{The search tree built by \PQE}
\vspace{18pt}
\label{fig:search_tree}
\end{wrapfigure}

\begin{wrapfigure}{l}{1.6in}
 \begin{center}
 \includegraphics[height=25mm]{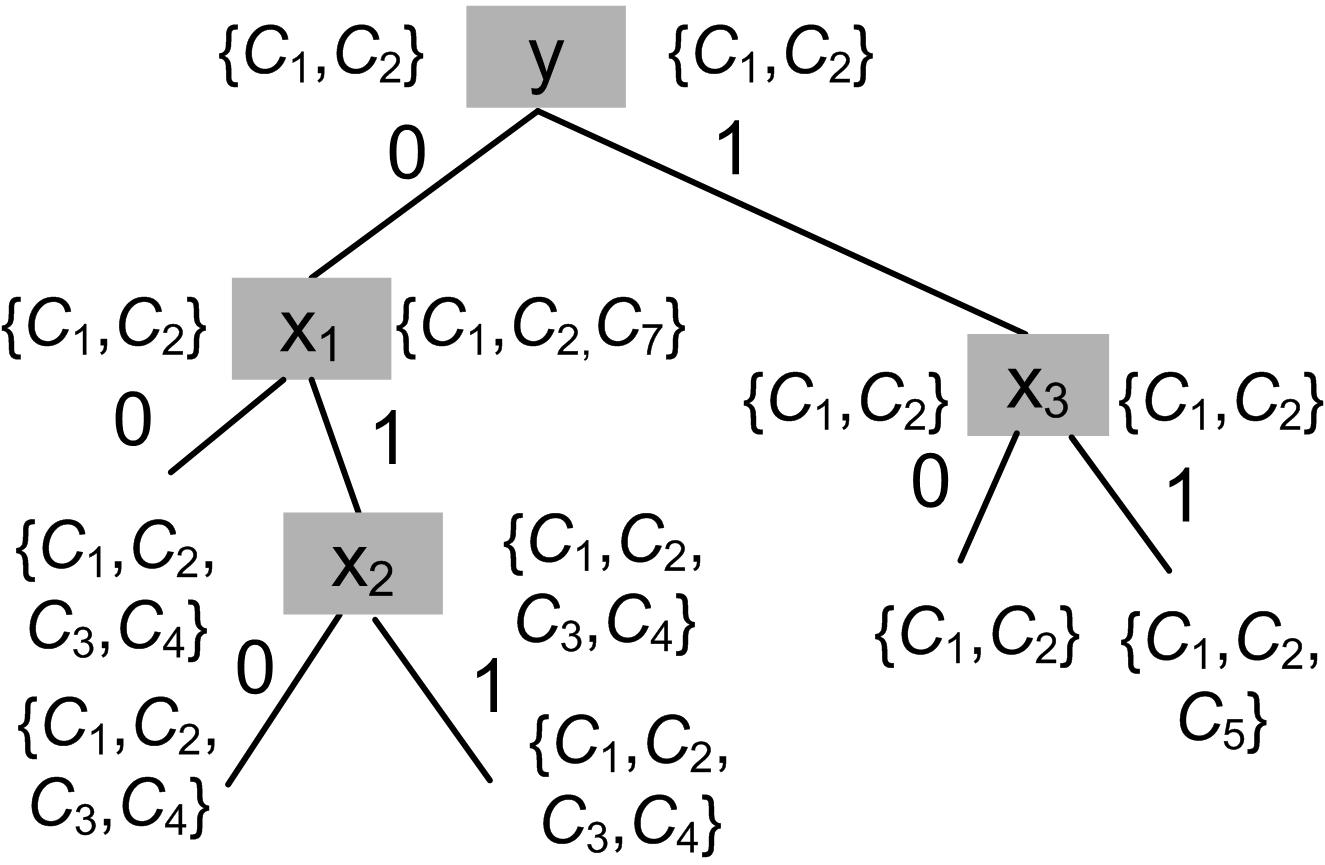}
  \end{center}
\caption{Dynamics of the \prf set}
\label{fig:pr_clauses}
\end{wrapfigure}

\ti{Search tree.} \PQE is a branching algorithm. It first proves
redundancy of $X$-clauses of $F$ in subspaces and then merges
results of different branches. When \PQE returns to the root of
the search tree, all the $X$-clauses of $F$ are proved redundant
in \prob{X}{F \wedge G}.  The search tree built by \PQE is given in
Figure~\ref{fig:search_tree}.  It also shows the nodes where new
clauses $C_7$ and $C_8$ were derived. \PQE assigns free
variables \ti{before} quantified.  For that reason, variable $y$
is assigned first.  At every node of the search tree specified by
assignment \pnt{q}, \PQE maintains a set of clauses denoted as
\prf\!\!. Here PR stands for ``clauses to Prove Redundant''. We
will refer to a clause of \prf as a \tb{PR-clause}.  \prf
includes all $X$-clauses of $F$ plus some $X$-clauses of $G$. The
latter are proved redundant to make proving redundancy of
$X$-clauses of $F$ easier. Sets \prf are shown in
Figure~\ref{fig:pr_clauses}. For every non-leaf node of the
search tree two sets of PR-clauses are shown. The set on the left
side (respectively right side) of node \pnt{q} gives \prf when
visiting node \pnt{q} for the first time (respectively when
backtracking to the right branch of node \pnt{q}).

\begin{wrapfigure}{l}{1.5in}
 \begin{center}
 \includegraphics[height=45mm]{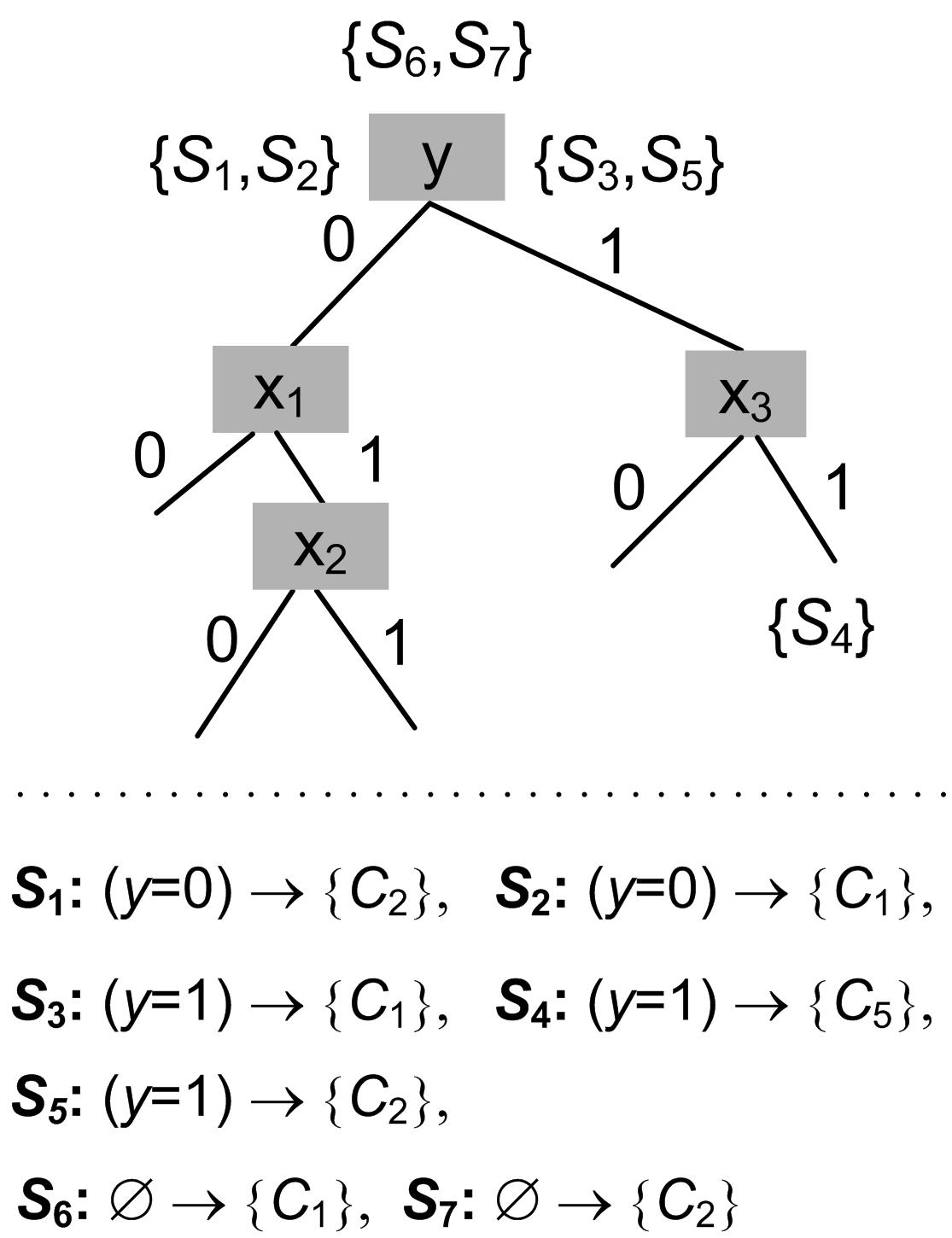}
  \end{center}
\caption{Derived D-sequents}
\label{fig:dseqs}
\end{wrapfigure}

\ti{Using D-sequents.} The main concern of \PQE is to prove
redundancy of PR-clauses. Branching is used to reach subspaces
where proving redundancy is easy.  The redundancy of a PR-clause
$C$ is expressed by a Dependency Sequent D-sequent. In short
notation, a D-sequent is a record \Ds{s}{C} saying that clause
$C$ is redundant in formula \prob{X}{F \wedge G} in any subspace
where assignment \pnt{s} is made. We will refer to \pnt{s} as the
\tb{conditional part} of the D-sequent.  The D-sequents
$S_1,\dots,S_7$ derived by \PQE are shown in
Figure~\ref{fig:dseqs}. They are numbered in the order they were
generated.  So-called atomic D-sequents record trivial cases of
redundancy.  More complex D-sequents are derived by a
resolution-like operation called \ti{join}. When \PQE returns to
the root, it derives D-sequents stating the unconditional
redundancy of the $X$-clauses of $F$.

\ti{Merging results of different branches.} Let $v$ be the
current branching variable and $v=0$ be the first branch explored
by \pqe.  After completing this branch, \PQE proves redundancy of
all clauses that currently have the PR-status. (The only
exception is the case when a PR-clause gets falsified in branch
$v=0$.  We discuss this exception below.)  Then \PQE explores
branch $v=1$ and derives D-sequents stating redundancy of clauses
in this branch.  Before backtracking from node $v$, \PQE uses
operation \ti{join} to produce D-sequents whose conditional part
does not depend on $v$.  For example, in branch $y=0$, D-sequent
$S_1$ equal to $(y=0) \rightarrow \s{C_2}$ was derived.  In
branch $y=1$, D-sequent $S_5$ equal to $(y=1) \rightarrow
\s{C_2}$ was derived. By joining $S_1$ and $S_5$ at variable $y$,
D-sequent $S_7$ equal to $\emptyset \rightarrow \s{C_2}$ was
produced where the conditional part did not depend on $y$.

\ti{Derivation of new clauses.} Note that redundancy of the
PR-clauses in subspace $y=1$ was proved without adding any new
clauses.  On the other hand, proving redundancy of PR-clauses in
subspace $y=0$ required derivation of clauses $C_7 =
\overline{x}_1$ and $C_8 = y$. For instance, clause $C_7$ was
generated at node $(y=0,x_1=1)$ by resolving $C_3$ and
$C_4$. Clause $C_7$ was \ti{temporarily} added to $F$ to make
PR-clauses $C_3$ and $C_4$ redundant at the node above.  However,
$C_7$ was removed from formula $F$ after derivation of clause
$C_8$ because the former is subsumed by the latter in subspace
$y=0$.  This is similar to conflict clause generation in
SAT-solvers where the intermediate resolvents are discarded.

\ti{Derivation of atomic D-sequents.} $S_1,\dots,S_5$ are the
atomic D-sequents derived by \pqe.  They record trivial cases of
redundancy. (Due to the simplicity of this example, the
conditional part of all atomic D-sequents has only assignment to
$y$ \ie the free variable. In general, however, the
conditional part of a D-sequent also contains assignments to
quantified variables.) There are three kinds of atomic
D-sequents. \ti{D-sequents of the first kind} state redundancy of
clauses satisfied in a subspace. For instance, D-sequent $S_1$
states redundancy of clause $C_2$ satisfied by assignment $y=0$.
\ti{D-sequents of the second kind} record the fact that a clause
is redundant because some other clause is falsified in the
current subspace.  For instance, D-sequent $S_2$ states that
$C_1$ is redundant because clause $C_8 = y$ is falsified in
subspace $y=0$.  \ti{D-sequents of the third kind} record the
fact that a clause is redundant in a subspace because it is blocked at
a variable $v$. That is this clause cannot be resolved on
$v$. For example, D-sequent $S_4$ states redundancy of $C_5$ that
cannot be resolved on $x_4$ in subspace $(y=1,x_3=1)$. Clause
$C_5$ is resolvable on $x_4$ only with $C_6$ but $C_6$ is
satisfied by assignment $y=1$.

\ti{Computation of the set of PR-clauses.}  The original set of
PR-clauses is equal to the the initial set of $X$-clauses of
$F$. Denote this set as \pri\!. In our example, \pri =
\s{C_1,C_2}. There are two situations where \prf is extended. The
first situation occurs when a parent clause of a new resolvent is
in \prf and this resolvent is an $X$-clause.  Then this resolvent
is added to \prf\!\!. An example of that is clause
$C_7=\overline{x_1}$ obtained by resolving PR-clauses $C_3$ and
$C_4$.

The second situation occurs when a PR-clause becomes unit.
Suppose a PR-clause $C$ is unit at node \pnt{q}, $v$ is the
unassigned variable of $C$ and $v \in X$.  \PQE first makes the
assignment falsifying $C$. Suppose that this is assignment
$v=0$. Note that all PR-clauses but $C$ itself are obviously
redundant at node $\pnt{q}~\cup (v=0)$.  \PQE backtracks and
explores the branch $v=1$ where clause $C$ is satisfied. At this
point \PQE extends the set $\mi{PR}(\pnt{q} \cup (v=1))$ by
adding \ti{every} clause of $F \wedge G$ that a) has literal
$\overline{v}$; b) is not satisfied; c) is not already
in \prf\!\!.

The extension of the set of PR-clauses above is done to guarantee
that clause $C$ will be proved redundant when backtracking off
the node \pnt{q}.  Let us consider the two possible cases. The
first case is that formula $F \wedge G$ is \ti{unsatisfiable} in
branch $v=1$. Then extension of the set of PR-clauses above
guarantees that a clause falsified by $\pnt{q} \cup (v=1)$ will
be derived to make the new PR-clauses redundant. Most
importantly, this clause will be resolved with $C$ on $v$ to
produce a clause rendering $C$ redundant in subspace \pnt{q}.
The second case is that formula $F \wedge G$ is \ti{satisfiable}
in branch $v=0$.  Then the redundancy of the clauses with literal $\overline{v}$ will be
proved without derivation of a clause falsified by $\pnt{q} \cup
(v=1)$.  When backtracking to node \pnt{q}, clause $C$ will be
blocked at variable $v$ and hence redundant. Note that extension
of the set \prf is \ti{temporary}. When \PQE backtracks past node
\pnt{q}, the clauses that became PR-clauses there lose their
PR-status.

Let us get back to our example. The first case above occurs at
node $y=0$ where PR-clause $C_1$ becomes unit.  \PQE falsifies
$C_1$ in branch $x_1=0$, backtracks and explores branch
$x_1=1$. In this branch, clauses $C_3,C_4$ of $G$ are made
PR-clauses. This branch is unsatisfiable. Making $C_3$,$C_4$
PR-clauses forces \PQE to derive $C_7 = \overline{x_1}$ that
makes $C_3,C_4$ redundant. But the real goal of obtaining $C_7$
is to resolve it with $C_1$ to produce clause $C_8 = y$ that
makes $C_1$ redundant.

The second case above occurs at node $y=1$ where clause $C_2$
becomes unit. Clause $C_2$ gets falsified in branch $x_3=0$. Then
\PQE backtracks and explores branch $x_3=1$. In this branch,
$C_5$ of $G$ becomes a new PR-clause as containing literal
$\overline{x}_3$. This branch is satisfiable and $C_5$ is proved
redundant without adding new clauses. Clause $C_2$ gets blocked
at node $y=1$ and hence redundant.

\ti{Forming a solution to the PQE problem.}  The D-sequents
derived by \PQE at a node of the search tree are \ti{composable}. This
means that the clauses that are redundant individually
are also redundant together. For example, on returning to the
root node, D-sequents $S_6$ and $S_7$ equal to $\emptyset \rightarrow \s{C_1}$
and $\emptyset \rightarrow \s{C_2}$ respectively are
derived. The composability of $S_6$ and $S_7$ means that
D-sequent $\emptyset \rightarrow \s{C_1,C_2}$ holds as well.  The
only new clause added to $F$ is $C_8 = y$ (clause $C_7$ was added
temporarily).  After dropping the $X$-clauses $C_1,C_2$ from $F$
as proved redundant one concludes that $y \wedge \prob{X}{G}
\equiv \prob{X}{F \wedge G}$ and $F^*= y$ is a solution to the
PQE problem.

\section{Basic Definitions}
\label{sec:basic_defs}
In this section, we give relevant  definitions.
%
%
\begin{definition}
\label{def:ecnf}
An \mbox{\boldmath{$\exists\mi{CNF}~~\mi{formula}$}} is a formula of the form
$\exists X [F]$ where $F$ is a Boolean CNF formula, and $X$ is a set of
Boolean variables.
 Let \pnt{q} be an
assignment, $F$ be a CNF formula, and $C$ be a clause. \Va{q}
denotes the variables assigned in \pnt{q}; \V{F} denotes the set
of variables of $F$; \V{C} denotes the variables of $C$;
and $\V{\exists X [F]} = \V{F} \setminus X$.
\end{definition}

We consider \ti{true} and \ti{false}  as a special kind of clauses.

%
%
\begin{definition}
\label{def:cofactor}
Let $C$ be a clause, $H$ be a CNF formula, and \pnt{q} be an 
assignment such that $\Va{q} \subseteq \V{H}$. Denote by \cof{C}{q} the clause equal to \emph{true} if $C$ is satisfied by
\pnt{q}; otherwise \cof{C}{q} is the clause obtained from $C$ by removing
all literals falsified by \pnt{q}.  \cof{H}{q} denotes the
formula obtained from $H$ by replacing every clause $C$ of $H$ with \cof{C}{q}. In this paper, we assume  
that  clause \cof{C}{q} equal to \ti{true} remains in \cof{H}{q}.
We treat such a clause as \ti{redundant} in \cof{H}{q}. Let \prob{X}{H} be an \ecnf~and \pnt{y} be an assignment to $\V{H} \setminus X$.
Then \cof{(\prob{X}{H})}{y} = \prob{X}{\cof{H}{y}}.
\end{definition}
%
%
\begin{definition}
\label{def:formula-equiv}
Let $S, Q$ be \ecnf~formulas. We say that $S, Q$ are \tb{equivalent},
written $S \equiv Q$, if for all assignments, $\pnt{y}$, such
that $\Va{y} \supseteq (\V{S} \cup \V{Q})$, we have $\cof{S}{y} =
\cof{Q}{y}$. Notice that $\cof{S}{y}$ and $\cof{Q}{y}$ have no
free variables, so by $\cof{S}{y} = \cof{Q}{y}$ we mean semantic
equivalence. 
\end{definition}
\vspace{4pt}
%
%
\begin{definition}
\label{def:qe-solution}
The \tb{Quantifier Elimination (QE) problem} for  \ecnf~formula
\prob{X}{H} is to find a CNF formula $H^*$ such that
$H^* \equiv \prob{X}{H}$. The \tb{Partial QE (PQE) problem} for \ecnf~formula
\prob{X}{F \wedge G} is to find a CNF formula $F^*$ such that
$F^* \wedge \prob{X}{G}  \equiv \prob{X}{F \wedge G}$.
\end{definition}

%
%
\begin{definition}
\label{def:red_vars}
Let $X$ be a set of Boolean variables, $H$ be a CNF formula and  $R$ be a subset of $X$-clauses of $H$.
The clauses of $R$ are 
\textbf{redundant} in CNF formula $H$ if 
$H \equiv (H \setminus R)$.
The clauses  of $R$ are 
\textbf{redundant} in \ecnf{} formula
$\exists X [H]$ if 
$\exists X [H] \equiv \exists X [H \setminus R]$. Note that  $H \equiv (H \setminus R)$
implies $\exists X [H] \equiv \exists X [H \setminus R]$ but the opposite is not true.
\end{definition}

\section{Dependency Sequents}
\label{sec:dseqs}
In this section, we recall   clause Dependency sequents (D-sequents) introduced in~\cite{fmcad13},
operation \ti{join} and the notion of composability.
In this paper, we will refer to clause D-sequents as just  D-sequents.

%
%
\begin{definition}
\label{def:d_sequent}
Let \prob{X}{H} be an \ecnf{} formula. Let \pnt{s} be an assignment to \V{H}
and $R$ be a subset of $X$-clauses of $H$.
A  dependency sequent (\tb{D-sequent})  has the form
\Dss{X}{H}{s}{R}. It states that 
the clauses of \cof{R}{s} are redundant in  \prob{X}{\cof{H}{s}}. 
Alternatively, we will  say that the clauses of $R$ are redundant in \prob{X}{H} in  subspace \pnt{s} (and in any other
subspace \pnt{q} such that  $\pnt{s} \subseteq \pnt{q}$).
\end{definition}

%
%
\begin{definition}
\label{def:res_part_assgns}
Let \pnt{s'} and \pnt{s''} be  assignments in which exactly  one variable $v \in \Va{s'} \cap \Va{s''}$ is assigned different values.
The assignment \pnt{s} consisting of all  the assignments of \pnt{s'} and \pnt{s''} but those to $v$ is called the
 \ti{resolvent} of \pnt{s'},\pnt{s''} on $v$.
Assignments \pnt{s'},\pnt{s''} are called \ti{resolvable}  on $v$.
\end{definition}

%
%
\begin{definition}
\label{def:join_rule}
Let \prob{X}{H} be an \ecnf{} formula. Let D-sequents
\Dss{X}{H}{s'}{R} and \Dss{X}{H}{s''}{R} hold. We refer to these
D-sequents as parent ones. Let \pnt{s'}, \pnt{s''} be resolvable on $v
\in \V{H}$ and \pnt{s} be the resolvent of \pnt{s'} and \pnt{s''}.  We
will say that D-sequent \Dss{X}{H}{s}{R} is obtained by \tb{joining}
the parents at $v$. The validity of this D-sequent is implied by that
of its parents ~\cite{fmcad13}.
\end{definition}

%
%
\begin{definition}
\label{def:consist_assignments}
Let \pnt{s'} and \pnt{s''} be assignments to a set of variables $Z$.
We will say that \pnt{s'} and \pnt{s''} are  \ti{compatible} if 
every variable of $\Va{s'} \cap \Va{s''}$ is assigned the same value
in \pnt{s'} and \pnt{s''}.
\end{definition}
%
%
\begin{definition}
\label{def:compose}
Let \Dss{X}{H}{s'}{R'} and  \Dss{X}{H}{s''}{R''} be two D-sequents
where \pnt{s'} and \pnt{s''} are compatible assignments to \V{H}.
We will call these D-sequents  \tb{composable} if
the D-sequent \Dss{X}{H}{s' \cup s''}{R' \cup R''} holds.
\end{definition}

\section{Algorithm}
\label{sec:algorithm}

In this section, we describe a PQE algorithm called \ti{\tb{DS-PQE}}
where DS stands for Dependency Sequents.  \PQE is based on our QE
algorithm \Cdi described in~\cite{fmcad13}. In this section, we will
mostly focus on the features of \PQE that differentiate it from
\Cdi\!\!. The algorithm description given in the first version of this
report~\cite{first_version} missed a case.  We address this case in
Subsections~\ref{ssec:big_picture} and ~\ref{ssec:new_features}.

%
%
%
%
\setlength{\intextsep}{2pt}
\setlength{\textfloatsep}{5pt}
\begin{figure}
\small
// \pnt{q} is an assignment to \V{F \wedge G} \\
// \DS~denotes a set of active D-sequents \\
// $\Phi$ denotes  \prob{X}{F \wedge G}\\
// $W$ denotes \prf \\
// If \Pqe returns  clause \ti{nil} (respectively a non-\ti{nil} clause), \\ 
//~~~~~\cof{(F \wedge G)}{q} is satisfiable (respectively unsatisfiable) \\
\vspace{-10pt}
\begin{tabbing}
aaaaa\=bb\= cc\= dddddddddddd\= \kill
\Pqe($\Phi$,$W$,\pnt{q},\DS)\{\\
\tb{\scriptsize{1}} \> if ($\exists$ clause $C \in F \cup G$ falsif. by \pnt{q}) \{\\ 
\tb{\scriptsize{2}}\Tt $\DS := \mi{atomic\_Dseqs1}(\DS,\pnt{q},C)$;\\  
\tb{\scriptsize{3}} \Tt   return($\Phi,\DS,C$);\}\\
\tb{\scriptsize{4}}\> $\DS := \mi{atomic\_Dseqs2}(\Phi,\pnt{q},\DS)$; \\
\tb{\scriptsize{5*}}\>if ($\mi{every\_PR\_clause\_redund}(W,\DS)$) return($\Phi,\DS,\mi{nil}$);\\
\> - - - - - - - - - - - -  \\
\tb{\scriptsize{6}}\> $v := \mi{pick\_variable}(F\wedge G,\pnt{q},\DS)$; \\
\tb{\scriptsize{7*}}\> $(\Phi,\DS,C_b):=$\Pqe\!\!($\Phi$,$W,\pnt{q}\cup(v=b)$,\DS);  \\
\tb{\scriptsize{8}}\> $\Sub{\DS}{asym}:= \mi{Dseqs\_to\_be\_inactive}(F,\DS,v)$;\\
\tb{\scriptsize{9}}\> if ($\Sub{\DS}{asym} = \emptyset$)  return($\Phi,\DS,C_b$);\\
\tb{\scriptsize{10}}\> $\DS := \DS \setminus \Sub{\DS}{asym};$ \\
\tb{\scriptsize{11*}}\> if $(\mi{impl\_assgn}(v,\overline{b}))$ $W' := \mi{newPRclauses}(W,F \wedge G,\overline{b})$; \\
\tb{\scriptsize{12*}}\> else $W' := \emptyset$; \\
\tb{\scriptsize{13*}}\> $(\Phi,\DS,C_{\overline{b}})$ := \Pqe($\Phi$,$W \cup W'$,$\pnt{q} \cup (v=\overline{b}$),\DS);\\
\> - - - - - - - - - - - - - \\ 
\tb{\scriptsize{14**}}\> if (($C_b =  \mi{nil}$) and ($C_{\overline{b}} \neq \mi{nil}$))\{\\
\tb{\scriptsize{15**}}\Tt $F := F \wedge C_{\overline{b}}$; \\
\tb{\scriptsize{16**}}\Tt $\DS := \mi{discard\_dseqs}(\DS,v)$; \\
\tb{\scriptsize{17**}}\Tt return($\Phi,\DS,\mi{nil}$);\} \\
\> - - - - - - - - - - - - - \\
\tb{\scriptsize{18}}\> if (($C_b \neq \mi{nil}$) and ($C_{\overline{b}} \neq \mi{nil}$))\{\\
\tb{\scriptsize{19}}\Tt  $C := resolve\_clauses(C_b,C_{\overline{b}},v)$; \\
\tb{\scriptsize{20}}\Tt  $F := F \wedge C$; \\
\tb{\scriptsize{21}}\Tt $\DS := \mi{atomic\_Dseqs1}(\DS,\pnt{q},C)$;\\  
\tb{\scriptsize{22*}}\Tt if (($C_b \in W$) or ($C_{\overline{b}} \in W$)) \\
\tb{\scriptsize{23*}}\ttt   $W := W \cup \s{C}$; \\
\tb{\scriptsize{24}}\Tt  return($\Phi,\DS,C$);\} \\
\tb{\scriptsize{25}}\> $\DS := \mi{merge}(\Phi,\pnt{q},v,\Sub{\DS}{asym},\DS,C_b,C_{\overline{b}})$;\\
\tb{\scriptsize{26}}\> return($\Phi,\DS,\mi{nil}$);\} \\
\end{tabbing} 
\vspace{-15pt}
\caption{\PQE procedure}
\vspace{10pt}
\label{fig:hl_descr}
\end{figure}

%

 \PQE derives D-sequents \DDs{X}{F \wedge G}{s}{C} stating the
 redundancy of $X$-clause $C$ in any subspace \pnt{q} such that
 $\pnt{s} \subseteq \pnt{q}$.  From now on, we will use a short
 notation of D-sequents writing \Ds{s}{C} instead of \DDs{X}{F \wedge
   G}{s}{C}.  We will assume that the parameter \prob{X}{F \wedge G}
 missing in \Ds{s}{C} is the \ti{current} \ecnf{} formula (with all
 resolvents added to $F$).  One can omit \prob{X}{F \wedge G} from
 D-sequents because \DDs{X}{F \wedge G}{s}{C} holds no matter how many
 resolvent clauses are added to $F$~\cite{fmcad13}. We will call
 D-sequent \Ds{s}{C} \tb{active} in subspace \pnt{q} if $\pnt{s}
 \subseteq \pnt{q}$. The fact that \Ds{s}{C} is active in subspace
 \pnt{q} means that $C$ is redundant in \prob{X}{F \wedge G} in
 subspace \pnt{q}.

%
%
\subsection{Input and output of \PQE}
Recall that a PR-clause is an $X$-clause of $F \wedge G$ whose
redundancy needs to be proved in subspace \pnt{q} (see
Section~\ref{sec:example}).  A description of \PQE is given in
Figure~\ref{fig:hl_descr}.  \PQE accepts an \ecnf{} formula \prob{X}{F
  \wedge G} (denoted as $\Phi$), an assignment \pnt{q} to \V{F}, the
set of PR-clauses (denoted as $W$) and a set \DS~of D-sequents active
in subspace \pnt{q} stating redundancy of \ti{some} PR-clauses in
\prob{X}{F \wedge G} in subspace \pnt{q}.

Similarly to Section~\ref{sec:example}, we will assume that the
resolvent clauses are added to formula $F$ while formula $G$ remains
unchanged.  \PQE returns a formula \prob{X}{F \wedge G} modified by
resolvent clauses added to $F$ (if any), a set \DS~of D-sequents
active in subspace \pnt{q} that state redundancy of \ti{all}
PR-clauses in \prob{X}{F \wedge G} in subspace \pnt{q} and a clause
$C$.  If \cof{(F \wedge G)}{q} is unsatisfiable then $C$ is a clause
of $F \wedge G$ falsified by \pnt{q}. Otherwise, $C$ is equal to
\ti{nil} meaning that no clause implied by $F \wedge G$ is falsified
by \pnt{q}.

The active D-sequents derived by \PQE are composable. That is if
$\Ds{s_1}{C_1},\ldots,\Ds{s_k}{C_k}$ are the active D-sequents of
subspace \pnt{q}, then the D-sequent \Ds{s^*}{C_1,\ldots,C_k} holds
where $\pnt{s^*} = \pnt{s_1} \cup \ldots \cup \pnt{s_k}$ and
$\pnt{s^*} \subseteq \pnt{q}$. Like \Cdi\!, \PQE achieves
composability of D-sequents by proving redundancy of PR-clauses in a
particular order (that can be different for different paths). This
guarantees that no circular reasoning is possible and hence the
D-sequents derived at a node of the search tree are composable.

 A solution to the PQE problem in subspace \pnt{q} is obtained by
 discarding the PR-clauses of subspace \pnt{q} (specified by $W$) from
 the CNF formula $F$ returned by \pqe.  To solve the original problem
 of taking $F$ out of the scope of the quantifiers in \prob{X}{F
   \wedge G}, one needs to call \PQE with $\pnt{q} = \emptyset$, $\DS
 = \emptyset,W = \pri$\!. Recall that \pri is the set of $X$-clauses
 of the original formula $F$.
%
%
\subsection{The big picture}
\label{ssec:big_picture}
\PQE consists of four parts separated in Figure~\ref{fig:hl_descr} by
the dotted lines.  In the first part (lines 1-5), \PQE builds atomic
D-sequents recording trivial cases of redundancy of $X$-clauses.  If
all the PR-clauses are proved redundant in \prob{X}{F \wedge G} in
subspace \pnt{q}, \PQE terminates at node \pnt{q}.

If some PR-clauses are not proved redundant yet, \PQE enters the
second part of the code (lines 6-13).  First, \PQE picks a branching
variable $v$ (line 6).  Then it recursively calls itself (line 7)
starting the left branch of $v$ by adding to \pnt{q} assignment $v=b$,
$b \in \s{0,1}$.  Once the left branch is finished, \PQE explores the
right branch $v=\overline{b}$ (line 13). 

The third part of \PQE (lines 14-17) takes care of the situation where
\begin{itemize}
\item  the left branch is satisfiable
\item the right branch is unsatisfiable
\item assignment $v = \overline{b}$ was not derived from a unit
  clause
\end{itemize}. 
(This situation was not mentioned in the first version of this
report~\cite{first_version}.)  In this case, \PQE simply
\begin{itemize}
\item adds to formula $F$ clause $C_{\overline{b}}$ derived in the
  right branch (and falsified by $\pnt{q} \cup (v = \overline{b})$),
\item discards D-sequents whose conditional part contains an
  assignment to variable $v$ (derived in the left and right branches)
  and backtracks.
\end{itemize}
Note that after backtracking, value $\overline{b}$ is derived from
$C_{\overline{b}}$ and the third part of \PQE is not invoked again
when branching on variable $v$. The reason for such a behavior of \PQE
is explained in Subsection~\ref{ssec:new_features}.

In the fourth part, \PQE merges the left and right branches (lines
18-26). This merging results in proving all PR-clauses redundant in
\prob{X}{F \wedge G} in subspace \pnt{q}.  For every PR-clause $C$
proved redundant in subspace \pnt{q}, the set \DS~contains precisely
one active D-sequent \Ds{s}{C} where $\pnt{s} \subseteq \pnt{q}$. As
soon as $C$ is proved redundant, it is marked and ignored until \PQE
enters a subspace \pnt{q'} where $\pnt{s} \not\subseteq \pnt{q'}$ \ie
a subspace where D-sequent \Ds{s}{C} becomes inactive. Then clause $C$
gets unmarked signaling that \PQE does not have a proof of redundancy
of $C$ in subspace \pnt{q'} yet.
\subsection{New features of \PQE with respect to \Cdi}
\label{ssec:new_features}
In this paper, we omit the description of functions of
Figure~\ref{fig:hl_descr} that operate identically to those of
\Cdi\!\!. What these functions do can be understood from the example
of Section~\ref{sec:example}. If this is not enough, the detailed
description of these functions can be found in ~\cite{fmcad13}. In
this subsection, we focus on the part of \PQE that is different from
\Cdi\!\!.  The lines of code of this part are marked with asterisks in
Figure~\ref{fig:hl_descr}. Lines 14-17 are marked with double
asterisks to indicate that they are not present in the first version
of this report~\cite{first_version}.

The main difference between \PQE and \Cdi is that at every node
\pnt{q} of the search tree, \PQE maintains a set \prf of PR-clauses.
\prf contains all the $X$-clauses of $F$ and some $X$-clauses of $G$
(if any).  \PQE terminates its work at node \pnt{q} when all the
current PR-clauses are proved redundant (line 5).  In contrast to
\pqe, \Cdi terminates at node \pnt{q}, when \ti{all} $X$-clauses are
proved redundant. Line 7 is marked because \PQE uses an additional
parameter $W$ when recursively calling itself to start the left branch
of node \pnt{q}.  Here $W$ specifies the set of PR-clauses to prove
redundant in the left branch.

Lines 11-12 show how \prf\!\! is extended. As we discussed in
Section~\ref{sec:example}, this extension takes place when assignment
$v=\overline{b}$ satisfies a unit PR-clause $C$. In this case, the set
$W'$ of new PR-clauses is computed. It consists of all the $X$-clauses
that a) contain the literal of $v$ falsified by assignment $v=
\overline{b}$; b) are not PR-clauses and c) are not satisfied. As we
explained in Section~\ref{sec:example}, this is done to facilitate
proving redundancy of clause $C$ at node \pnt{q}. The set $W'$ is
added to $W$ before the right branch is explored (line 13). Notice
that the clauses of $W'$ have PR-status only in the subtree rooted at
node \pnt{q}.  Upon return to node \pnt{q} from the right branch, the
clauses of $W'$ lose their PR-status.

Lines 14-17 address the special situation described in
Subsection~\ref{ssec:big_picture}: the left branch is satisfiable and
clause $C_{\overline{b}}$ is derived in the right branch, the latter
being unsatisfiable.  The problem here is as follows.  To prove
redundancy of clause $C_{\overline{b}}$, one needs to show redundancy
of clauses that can be resolved with $C_{\overline{b}}$ on variable
$v$. The redundancy of such clauses is supposed to be proved in the
left branch. However, the left branch was examined when clause
$C_{\overline{b}}$ was not in formula $F$. So, \PQE could not compute
the set of PR-clauses of the left branch correctly. To solve this
problem, \PQE simply adds $C_{\overline{b}}$ to $F$ and backtracks
unassigning variable $v$. Note that now assignment $v = \overline{b}$
can be derived from clause $C_{\overline{b}}$. So \PQE knows that it
needs to prove redundancy of clauses that can be resolved with
$C_{\overline{b}}$ (line 11).

As we mentioned in Section~\ref{sec:example}, one more source of new
PR-clauses are resolvents (lines 22-23). Let $v=b$ and
$v=\overline{b}$ be unsatisfiable branches and $C_b$ and
$C_{\overline{b}}$ be the clauses returned by \PQE. If $C_b$ or
$C_{\overline{b}}$ is currently a PR-clause, the resolvent $C$ becomes
a new PR-clause. One can think of a PR-clause as supplied with a tag
indicating the level up to which this clause preserves its
PR-status. If only one of the clauses $C_b$ and $C_{\overline{b}}$ is
a PR-clause, then $C$ inherits the tag of this clause.  If both
parents have the PR-status, the resolvent inherits the tag of the
parent clause that preserves its PR-status longer.
\subsection{Correctness of \PQE}
The correctness of \PQE is proved similarly to that of
\Cdi\cite{fmcad13}.  \PQE is complete because it examines a finite
search tree.  Here is an informal explanation of why \PQE is sound.
First, the clauses added to $F$ are produced by resolution and so are
correct in the sense they are implied by $F \wedge G$. Second, the
atomic D-sequents built by \PQE are correct. Third, new D-sequents
produced by operation \ti{join} are correct. Fourth, the D-sequents of
individual clauses are composable.

So when \PQE returns to the root node of the search tree, it derives
the correct D-sequent $(\prob{X}{F \wedge G},\emptyset) \rightarrow
F^X$. Here $F^X$ denotes the set of all $X$-clauses of $F$. Thus, by
removing the $X$-clauses from $F$ one obtains formula $F^*$ such that
$\prob{X}{F^* \wedge G} \equiv \prob{X}{F \wedge G}$. Since $F^*$ does
not depend on variables of $X$ it can be taken out of the scope of
quantifiers.


\section{Background}
\label{sec:background} 
QE has been studied by many researchers, due to its important
role in verification \eg in model checking.  QE methods are
typically based on BDDs~\cite{bryant_bdds1,bdds_qe} or
SAT~\cite{blocking_clause,fabio,
  cofactoring,cav09,hvc,cav11,cmu}. At the same time, we do not
know of research where the PQE problem was solved or even
formulated. Of course, identification and removal of redundant
clauses is often used in preprocessing procedures of
QBF-algorithms and SAT-solvers~\cite{prepr,blocked_qbf}.
However, these procedures typically exploit only situations
where clause redundancies are obvious.
 

PQE is different from QE in at least two aspects. First, a
PQE-algorithm has to have a significant degree of
``structure-awareness'', since PQE is essentially based on the
notion of redundancy. So it is not clear, for example, if a
BDD-based algorithm would benefit from replacing QE with
PQE. This also applies to many SAT-based algorithms of QE.  For
instance, in~\cite{fmcad12} we presented a QE algorithm called
DDS that was arguably more structure aware than its SAT-based
predecessors.  DDS is based on the notion of D-sequents
defined in terms of variable redundancy.  DDS makes quantified
variables redundant in subspaces and merges the results of
different branches.  Despite its structure-awareness, it is hard
to adjust DDS to solving PQE: in PQE, one, in general, does not 
eliminate quantified variables (only some clauses with quantified variables
are eliminated).

The second interesting aspect of PQE is as follows. QE can be
solved by a trivial albeit inefficient algorithm. Namely, to find
a quantifier-free formula equivalent to \prob{X}{H} one can just
resolve out all variables of $X$ as it is done in the DP
procedure~\cite{dp}.  However, the PQE problem does not have a
counterpart of this algorithm \ie PQE does no have a ``trivial''
PQE-solver.  Let $C$ be a clause of $H$ and $v$ be a variable of
$C$.  One can always make $C$ redundant by adding to $H$ all
resolvents of $C$ with clauses of $H$ on
$v$~\cite{allen,inproc}. So one can always ``resolve out'' any
clause of a CNF formula.  It seems that one can take formula $F$
out of the scope of the quantifiers in \prob{X}{F \wedge G} using
the following procedure.  Keep resolving out clauses of $F$ and
their resolvents with $G$ until all non-redundant resolvents
depend only on free variables.  Unfortunately, this
procedure may loop \ie a previously seen set of clauses $F \wedge
G$ may be reproduced later. \PQE does not have this problem due
to branching.



\section{Experimental Results}
\label{sec:experiments}
Since we are not aware of another tool performing PQE, in the
experiments we focused on contrasting PQE and QE.  Namely, we compared
\PQE with our QE algorithm called \Cdi\cite{fmcad13}.  The fact that
\PQE and \Cdi are close in terms of implementation techniques is
beneficial: any difference in performance should be attributed to
difference in algorithms rather than implementations.  

In the experiments, we used \PQE and \Cdi for backward model checking.
We will refer to model checkers based on \PQE and \Cdi as \mcpq and
\mcq respectively. The difference between \mcpq and \mcq is as
follows. Let $F(S')$ and $T(S,S')$ specify a set of next-states and
transition relation respectively. The basic operation here is to find
the pre-image $H(S)$ of $F$ where $H \equiv \prob{S'}{F \wedge T}$. So
$H$ is a solution to the QE problem.  As we showed in
Subsection~\ref{subsec:pre-image}, one can also find $H$ just by
taking $F$ out of the scope of the quantifiers in formula \prob{S'}{F
  \wedge T}.  \mcq computes $H$ by making redundant \ti{all}
$S'$-clauses of $F \wedge T$ while \mcpq finds $H$ by making redundant
\ti{only} the $S'$-clauses of $F$.

\setlength{\intextsep}{4pt}
\begin{wrapfigure}{l}{2in}
 \begin{center}
\includegraphics[width=2in]{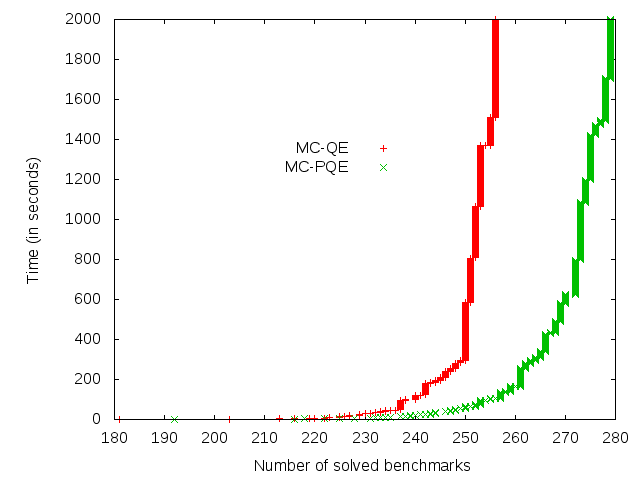}
  \end{center}
\caption{Performance of  model checkers on 282 examples solved by \mcq or \mcpq}
\vspace{5pt}
\label{fig:model_checking}
\end{wrapfigure}

The current implementations of \Cdi\! and \PQE lack D-sequent
re-using: the parent D-sequents are discarded after a join
operation. We believe that re-using D-sequents should boost
performance like clause recording in SAT-solving. However, when
working on a new version of \Cdi we found out that re-using
D-sequents indiscriminately may lead to circular reasoning.  We
have solved this problem theoretically and resumed our work on
the new version of \Cdi\!\!. However, here we report the
results of implementations that do not re-use D-sequents.

%
%
\vspace{5pt}
\begin{table}[htb]
\small
\caption{\ti{Model checking results on some concrete examples}}
\scriptsize
\begin{center}
\begin{tabular}{|p{42pt}|p{14pt}|l|p{14pt}|p{12pt}|p{18pt}|p{16pt}|} \hline
 benchmark & \#lat- & \#gates & \#ite- & bug  &{\tiny MC-} & {\tiny MC-} \\ 
           &   ches &         & rati-    &   & {\tiny QE}  & {\tiny PQE}  \\ 
          &         &         & ons   &     & (s.)  & (s.) \\ \hline
bj08amba3g62 &32  &9,825&4&no   &241& \tb{38}   \\ \hline
kenflashp03  & 51  &3,738 & 2 & no  & \tb{33} & 104   \\ \hline
pdtvishuffman2 & 55 &831 &6& yes &$>$2,000 & \tb{296} \\ \hline
pdtvisvsar05 & 82 &2,097&4&no&1,368 & \tb{7.7} \\ \hline
pdtvisvsa16a01 &188  & 6,162& 2 & no & $>$2,000 & \tb{17}\\ \hline
texaspimainp12 & 239 & 7,987&4&no&807& \tb{580} \\ \hline
texasparsesysp1 &312  & 11,860  & 10  & yes  & 39  & \tb{25}  \\ \hline
pj2002 & 1,175 & 15,384 & 3 & no  & 254 & \tb{47}  \\ \hline
mentorbm1and & 4,344 & 31,684 & 2 & no &\tb{1.4} & 1.7 \\ \hline
\end{tabular}                
\label{tbl:concr_exmps}
\end{center}
\end{table}

We compared \mcpq and \mcq on the 758 benchmarks of HWMCC-10
competition~\cite{hwmcc10}.  With the time limit of 2,000s, \mcq
and \mcpq solved 258 and 279 benchmarks respectively. On the set
of 253 benchmarks solved by both model checkers, \mcpq was about
2 times faster (the total time is 4,652s versus 8,528s). However,
on the set of 282 benchmarks solved by at least one model checker
\mcpq was about 6 times faster (10,652s versus 60,528s).  Here we
charged 2,000s, \ie the time limit, for every unsolved
benchmark.

Figure~\ref{fig:model_checking} gives the performance of \mcq and
\mcpq on the 282 benchmarks solved by at least one model checker in
terms of the number of problems finished in a given amount of time.
Figure~\ref{fig:model_checking} shows that \mcpq consistently
outperformed \mcq\!.  Model checking results on some concrete benchmarks
are given in Table~\ref{tbl:concr_exmps}.  The column \ti{iterations}
show the number of backward images computed by the algorithms before
finding a bug or reaching a fixed point.

\section*{Acknowledgment}
This research was supported in part by DARPA under AFRL
Cooperative Agreement No.~FA8750-10-2-0233 and by NSF grants
CCF-1117184 and CCF-1319580.

\section{Conclusion}
\label{sec:conclusions}
We introduced the Partial Quantifier Elimination problem (PQE), a
generalization of the Quantifier Elimination problem (QE). We
presented a PQE-algorithm based on the machinery of D-sequents
and gave experimental results showing that PQE can be much more
efficient than QE. Efficient PQE-solver may lead to new methods
of solving old problems like SAT-solving.  In addition, many
verification problems can be formulated and solved in terms of
PQE rather than QE, a topic ripe for further exploration.

\bibliographystyle{plain}
\bibliography{short_sat,local}
\end{document}